# One more (a further) generalization of the Final Value Theorem

*Emanuel Gluskin*[1,2,*], *Shmuel Miller*[2], *Joris Walraevens*[3]

[1]Kinneret College, Israel, [2]Braude College, Israel, [3]Ghent University, Belgium.
*Communicating author (gluskin@ee.bgu.ac.il)

**Abstract:** It is shown that the previous [1-3] generalization of the final-value theorem to the average (not necessarily limiting) values, can be extended to the higher-order running averages ($<>_t$), $\lim_{s\to 0}[sF(s)] = \lim_{t\to\infty} <<...<<f(\lambda_q)>_{\lambda_{q-1}}>_{\lambda_{q-2}}...>_{\lambda_1}>_t$, if *f*(*t*) is such that $\exists m = \min[q] < \infty$ for which the later limit exists.

## 1. Introduction

In [1-3], the final-value, Laplace and z-transform, theorems were generalized, respectively, to

$$\lim_{s\to 0}[sF(s)] = \lim_{t\to\infty} < f(\lambda) >_t = \lim_{t\to\infty} \frac{1}{t}\int_0^t f(\lambda)d\lambda \qquad (1)$$

and

$$\lim_{z\to 1}[(z-1)F(z)] = \lim_{n\to\infty} < f[k] >_n = \lim_{n\to\infty} \frac{1}{n}\sum_{k=0}^{n} f[k], \qquad (2)$$

that is with the replacement of the limiting value $f(\infty)$ of the function $f(t)$ (or the limiting value $f[\infty]$ of the sequence $f[n]$) at infinity, by the limit of its *running average* over (0,*t*) or (0,*n*), as *t* or *n* tending to infinity.

In terms of only continuous-time case, we extend here the generalization of the final value theorem, as follows.

Assume that $f(t)$, as well as the running "arithmetic" average over (0,*t*),

$$\psi_1(t) = < f >_t \equiv \frac{1}{t}\int_0^t f(\lambda)d\lambda,$$

have no limit as $t \to \infty$, but there exists some natural *m* for which the *m*-order running average

$$\psi_m(t) = \frac{1}{t}\int_0^t (\frac{1}{t_1}\int_0^{t_1} (\frac{1}{t_2}\int_0^{t_2} ...(\frac{1}{t_{m-1}}\int_0^{t_{m-1}} f(t_m)dt_m)\, dt_{m-1})...)dt_2)dt_1 \qquad (3)$$

that might be compactly denoted as $\psi_m(t) = <<...<<f(\lambda_m)>_{\lambda_{m-1}}>_{\lambda_{m-2}}...>_{\lambda_1}>_t$ has the limit,

$$\psi_m(\infty) = \lim_{t\to\infty} \psi_m(t).$$

Since from existence of $\lim \psi_m(t),\ t \to \infty$ the existence of all of the limits $\lim \psi_{m+n}(t),\ n = 1,2,...,t \to \infty,$ follows (smoothing improves the convergence), $m$ will be understood below as the *minimal* order of the "running averages", converging as $t \to \infty$.

It is worth stressing (especially for a teacher), that this is also the situation with the classical final-value theorem arising here for $m$ = 0, and also with the generalized theorem [1-3] for $m$ = 1.

Reserving thus the index $m$ for the minimal order of the averaging, we shall, in the more general consideration, label the sequential averages using index $q$ = 0,1,2,3, …, starting $\psi_q(t)$ from $\psi_o(t) = f(t)$ . For instance

$$\psi_2(t) = << f >_\lambda >_t \equiv \frac{1}{t}\int_0^t \left( \frac{1}{\lambda} \int_0^\lambda f(\mu) d\mu \right) d\lambda \ . \qquad (4)$$

We first consider some simple relevant functions that show that the situation with the existence of the averages is, generally, not trivial, and then prove the central equation (18) whose sense is that all $\psi_q(t)$ of different order ($q \geq m$), belonging to the same *f*(*t*), should have the same asymptotic behavior (see also [1,2]) as $t \to \infty$.

## 2. Example of *f*(*t*) with a finite *m* > 1

Consider the function

$$f_p(t) \ = \ t^p \sin t \ \ (\text{or,} \ \ t^p \cos t\,), \qquad (5)$$

where $p > 0$ is integer.

Such functions as (5) are of interest here because for $p \geq 0$ we have (see Lemma 2 below) the nontrivial case of $m = p+1 \geq 1$, and because in view of the linearity of averaging, study of such functions (Lemma 3) provides an introduction to study the more general functions $t^p \varphi(t)$ where $\varphi(t)$ is either periodic or almost periodic.

The average $\psi_q(t)$ associated with (5) will be denoted as $\psi_q(t,p)$, and, correspondingly, $\psi_m(t)$ as $\psi_m(t,p)$.

Consider the first averaging, i.e. the following equality that is correct for $p \geq 1$,

$$\begin{aligned}\psi_1(t,p) = < \mu^p \sin\mu >_t &= \frac{1}{t}\int_0^t \lambda^p \sin\lambda \, d\lambda \\ &= -t^{p-1}\cos t + \frac{p}{t}\int_0^t \lambda^{p-1}\cos\lambda \, d\lambda \qquad (6) \\ &= -t^{p-1}\cos t + p < \mu^{p-1}\cos\mu >_t, \quad p \geq 1.\end{aligned}$$

which for $p \geq 2$ can be also rewritten as:

$$<\mu^p \sin\mu>_t = t^{p-2}(p\sin t - t\cos t) - p(p-1)\frac{1}{t}\int_0^t \lambda^{p-2}\sin\lambda\, d\lambda \quad , \qquad (7)$$
$$= t^{p-2}(p\sin t - t\cos t) - p(p-1) <\mu^{p-2}\sin\mu>_t .$$

For $p$ =1, (6) becomes:

$$\psi_1(t,1) = <\mu\sin\mu>_t = \frac{\sin t}{t} - cos t . \qquad (7b) \quad (8)$$

One sees that for $q$ = 1 and any $p$, there are non-converging as $t \to \infty$ oscillating terms For $p$ = 1 there is one such term $-cos t$, and $\psi_1(\infty,1)$ does not exist.

However, for $p$ = 1, in the iterated averaging $\psi_2(t,1)$ the non-converging term is eliminated as $t \to \infty$:

$$\psi_2(t,1) = << \mu\sin\mu>_\lambda>_t = \frac{1}{t}\int_0^t \frac{\sin\lambda}{\lambda} d\lambda - \frac{1}{t}\int_0^t cos\lambda\, d\lambda ,$$

and

$$\psi_2(\infty,1) = \lim_{t\to\infty}\psi_2(t,1) = \lim_{t\to\infty}\left(\frac{1}{t}\int_0^t \frac{\sin\lambda}{\lambda} d\lambda - \frac{1}{t}\int_0^t cos\lambda\, d\lambda\right) = 0 \ .$$

In this particular case, not only the classical final value theorem related to $q$ = 0, also the generalization of [1] related to $q$ = 1, are insufficient, and we need the iterated averaging.

Observe that for the not oscillating $f(t) = t^p$, we have $\psi_1(t) \sim \frac{t^{p+1}}{t} = t^p$, etc., i.e. no $m$ exists.

<u>Remark</u>: Observe that the physical unit of any $\psi_q(t)$ is the same as that of *f*(*t*). This does not contradict the reduction in the degree of $t^p$ (i.e. *p* → *p-1*) with each successive averaging, thus emphasizing the importance of the dimensional (e.g. time-scaling) considerations. One, concerned with the physical dimensions of the functions involved, can use that for any interval of averaging,

$$<\mu^p \sin a\mu>_t = (a\mu = \tau) = \frac{1}{a^p} <\tau^p \sin\tau>_{at}, \quad .$$

That is, with the replacement $t \to at$ made at each place where *t* arises, factor $a^{-p}$ should be added before $\psi_m(at)$. Thus, for instance, the expression $<\mu^{p-2}\sin\mu>_t$ appearing in (7), becomes $a^{-(p+1)} <(a\mu)^{p-2}\sin a\mu>_{at}$. □

Comparison with $\lim_{s\to 0} sF(s)$ will be done in the next section.

## 3. More about the averaging of $f_p(t) = t^p \sin(t)$

Lemma 2: *For* $f_p(t) = t^p \sin t$ (similarly for, $t^p \cos t$), $m = p+1$, that is, we need *p+1* iterated averaging in order to come to existing limit as $t \to \infty$; then

$$\psi_{m+n}(\infty) = \psi_{p+1+n}(\infty) = 0, \quad n = 0,1,2, \ldots .$$

Proof: According to (7), the first average

$$\psi_1(t) = \frac{1}{t}\int_0^t \lambda^p \sin\lambda \, d\lambda$$

includes (in the term $t^{p-1}\cos t$) its highest degree of $t$ as $t^{p-1}$, and each next averaging further reduces the degree by 1. Since the limit as $t \to \infty$ exists only when we already come to $t^{-1}\cos t$ (since $t^{-1}\cos t \to 0,\ t \to \infty$), we have to require that $p - m = -1$, i.e. that $m = p+1$. □

As for the $s$-domain, using that [4] for $f(t) = t^p \sin\omega t$,

$$F(s) = \frac{\Gamma(p+1)[(s+j\omega)^{p+1} - (s-j\omega)^{p+1}]}{2j(s^2+\omega^2)^{p+1}} , \qquad (9)$$

we have for $q$ = 1 and *any p*

$$\lim_{s\to 0} sF(s) = 0 . \qquad (9a)$$

Note from (9) that for $p$ even, $sF(s) \sim s$, and for $p$ odd, $sF(s) \sim s^2$, as $s \to 0$.

Thus, the equality

$$\lim_{s\to 0} sF(s) = \psi_{p+1+n}(\infty) = 0, \quad n = 0,1,2, \ldots ,$$

is correct, though the role of $p$ can be seen in the $s$-domain only via the mentioned asymptotic expressions of $sF(s)$, as $s \to 0$.

This general conclusion agrees with the above example of $q$ = 1 and $p$ = 1. (Namely, $\lim_{s\to 0} sF(s)$ equals $\psi_2(\infty,1)$, but not $\psi_1(\infty,1)$, the latter does not exist).

The conclusions for $f(t) = t^p \cos\omega t$ are similar.

## 4. The case of $f_p(t) = t^p \varphi(t)$ with a periodic or almost-periodic $\varphi(t)$

The consideration of the *s*-domain is extended (for $q = 1$) by the following Lemma, to the functions in which the sinusoidal factor is replaced by an arbitrary periodical or almost periodic function.

Lemma 3:

Consider

$$f_p(t) = t^p \varphi(t) \tag{10}$$

where

$$p \in \mathbb{N}, \quad \varphi(t+T) = \varphi(t), \quad t \geq 0, \quad T > 0.$$

Obviously,

$$F_p(s) = (-1)^p \frac{d^p \Phi(s)}{ds^p}, \; p \geq 0 \; . \tag{11}$$

Expending

$$\varphi(t) = \sum_{k=-\infty}^{\infty} c_k e^{jk\omega t}, \quad \omega = \frac{2\pi}{T}, \; j = \sqrt{-1}, \tag{12}$$

we have

$$\Phi(s) = \sum_k \frac{c_k}{s - jk\omega}, \tag{13}$$

thus obtaining from (11):

$$F_p(s) = p! \sum_k \frac{c_k}{(s - jk\omega)^{p+1}} \; . \tag{14}$$

Obviously,

$$\lim_{s \to 0} [sF_p(s)] \tag{15}$$

is contributed only by the term with $k = 0$, and we obtain this limit as

$$\begin{matrix} \infty, \; if \;\; c_o \neq 0, \; p > 0. \\ 0, \; if \;\; c_o = 0. \end{matrix} \tag{16}$$

*independently of p*.

This agrees with (9)-(9a).

Furthermore, for $p = 0$, and $c_o \neq 0$, we get from (14)

$$\lim_{s \to 0} [sF_p(s)] = c_o . \tag{17}$$

This limit says, however, nothing about the value of $q$ for which

$$\lim_{t \to \infty} \psi_q(t) = c_o .$$

The linearity of all the expressions involved allows us to continue using the somewhat more general almost-periodic functions denoted as $\varphi_{a.p.}(t)$,

$$\varphi_{a.p.}(t) = \sum_k c_k e^{\lambda_k t} \tag{18}$$

with some real-valued { $\lambda_k$ } (and { $c_k$ }), and the use of (17) allows us to further explore the Laplace-domain part of the final-value theorem.

For such a function (see (14) ),

$$\psi_1(t) = \frac{1}{t}\int_0^t f_p(\mu)\,d\mu = \frac{1}{t}\int_0^t \mu^p \varphi_{a.p.}(\mu)\,d\mu \ .$$

A Laplace transform of $\psi_1(t)$ yields

$$\Psi_1(s) = \int_s \frac{1}{\zeta} F_p(\zeta)\,d\zeta = p!\int_s \sum_k \frac{c_k}{\zeta\cdot(\zeta - j\lambda_k)^{p+1}}\,d\zeta \ .$$

and

$$\lim_{s\to 0}[s\Psi_1(s)] = p!\lim_{s\to 0}[\,s\int_s \sum_k \frac{c_k}{\zeta\cdot(\zeta - j\lambda_k)^{p+1}}\,d\zeta]$$

$$= p!\lim_{s\to 0}[\frac{\int_s \sum_k \frac{c_k}{\zeta\cdot(\zeta - j\lambda_k)^{p+1}}\,d\zeta}{\frac{1}{s}}] =$$

$$= p!\lim_{s\to 0}[\frac{-\sum_k \frac{c_k}{s\cdot(s - j\lambda_k)^{p+1}}}{-\frac{1}{s^2}}]$$

$$= p!\lim_{s\to 0}[\,s\sum_k \frac{c_k}{(s - j\lambda_k)^{p+1}}] \ .$$

where L'Hospital rule was used.

Asymptotically,

$$s\sum_{k=0}^{\infty} \frac{c_k}{(s - jk\omega)^{p+1}} \underset{s\to 0}{\sim} \frac{c_o}{s^p} + s\sum_{k=1}^{\infty} \frac{c_k}{(-jk\omega)^{p+1}} \tag{18 ???}$$

which for a small $s$ , and $c_o \neq 0$ is $\sim \frac{c_o}{s^p}$.

Thus, for $c_o \neq 0$

$$s\Psi_1(s) \sim p!\frac{c_o}{s^p}$$

or

$$\Psi_1(s) \sim c_o \frac{p!}{s^{p+1}}, \quad \text{as } s \to 0 .$$

No limit exists, and for the formally respective time-function

$$\psi_1(t) \sim c_o t^p, \quad \text{as } t \to \infty$$

also there is no limit.

If $c_o \neq 0$, then in (18)

$$s \sum_{k=0}^{\infty} \frac{c_k}{(s - jk\omega)^{p+1}} \underset{s \to 0}{\sim} K s ,$$

where the constant $K$ is

$$K = \left( \sum_{k=1}^{\infty} \frac{c_k}{(-jk\omega)^{p+1}} \right)$$

The function $sF(s) = Ks$ corresponds in time domain to the *derivative* of $K\delta(t)$ including Dirac's delta function:

$$K\delta'(t) .$$

Time average of this function is zero, obviously.

Remark: Note that $f_p(t) = t^p \varphi(t)$ includes the case of the essentially positive functions as

$$f_p(t) = t^p \, |\sin t|, \quad f_p(t) = t^p \sin^2 t \ . \ \square$$

## 5. Possible physical origin of the functions $f_p(t) = t^p \sin t$ and $f_p(t) = t^p \varphi(t)$

For seeking applications of the function $f_p(t) = t^p \sin t$ that seems at this stage to be most promising from the mathematical side, one should consider an LTI homogeneous system (equation) with real coefficients, having two $p$-multiple eigenvalues $j$ and $-j$. Such an equation, which can be

$$(D^2 + 1)^p y(t) = 0, \quad D = \mathrm{d}/\mathrm{d}t , \qquad (19)$$

possesses solution of the type

$$f_p(t) = \sum_{\gamma=0}^{p} k_\gamma t^\gamma \sin t \qquad (20)$$

with some constants $k_\gamma$, $\gamma = 0,1,...,p$.

Using proper initial conditions, we can make $k_\gamma = 0$ for $\gamma = 0,1,...,p$ -1, having

$$f_p(t) = k_p t^p \sin t \,.$$

Realization of $f_p(t) = t^p \varphi(t)$ including the spectrally more complicated $\varphi(t)$ is obtained as a particular case, by the respectively more complicated LTI system having $p_1$ pairs of some conjugated complex eigenvalues ($\lambda_1, \lambda_1^*$), $p_2$ pairs of ($\lambda_2, \lambda_2^*$), etc., with all $\lambda_1, \lambda_2, \lambda_3 ...$ imaginary (that is, $p_1$ pairs of the zeroes $(j|\lambda_1|, -j|\lambda_1|)$; $p_2$ times of $(j|\lambda_2|, -j|\lambda_2|)$, etc..) .

Equations of such a system, of a sufficiently high total degree, which allow us to compose, similarly using their solutions, $y_1, y_2, ...$, the function $f_p(t) = t^p \varphi(t)$, can be

$$\begin{aligned} &(D^2 + \lambda_1^2)^{p_1} y_1(t) = 0 \\ &(D^2 + \lambda_2^2)^{p_2} y_2(t) = 0 \\ &\ldots\ldots \end{aligned} \qquad (21)$$

as many as needed.

Taking solutions of any such particular systems with the same *p* everywhere, but with the different $\{\lambda\}$, we can obtain the linear combination, $f_p(t) = t^p \varphi(t)$ .

## 6. The central equality: $\lim_{s\to 0} sF(s) = \lim_{t\to\infty} \psi_m(t) = \psi_m(\infty)$

Since the physical units (dimension) of each $\psi_m(t)$ are just those of *f*(*t*), *i.e. those of* $sF(s)$, the dimensional argument suggests clarifying whether or not, if $\exists m < \infty$, it may simply be not just for $m = 0$, or 1, that for $\forall q \geq m$

$$\lim_{s\to 0} sF(s) = \lim_{t\to\infty} \psi_q(t) = \lim_{t\to\infty} \psi_m(t) = \psi_m(\infty) \,. \qquad (22)$$

Theorem 1:

*If* $\exists m < \infty$, *then all of the* $\psi_{m+n}(t)$, $n = 1,2,3,...$, (i.e. $q = m+n$) *have the same asymptotic behavior as* $t \to \infty$. (We allow here tendency to zero, as a case of the asymptotic behavior)

We just have to prove (22) for $n = 0$, i.e. for $q = m$, because the equality

$$\lim_{s\to 0} [sF(s)] = \psi_{m+n}(\infty), \quad n > 0 \,, \qquad (23)$$

immediately follows then from the fact that further smoothing "$<>_t$" can only improve the convergence.

Proof of Theorem 1:

Using the correspondence

$$\psi_1(t) = \langle f \rangle_t = \frac{1}{t}\int_0^t f(\lambda)\,d\lambda \;\leftrightarrow\; \int_s^\infty \frac{F(\varsigma)}{\varsigma}\,d\varsigma\,, \tag{24}$$

we have, by simple repetitions of the operators involved in (24), that

$$\begin{aligned}\psi_m(t) &= \frac{1}{t}\int_0^t \frac{dt_1}{t_1}\int_0^{t_1} \frac{dt_2}{t_2}\int_0^{t_2} \ldots \int_0^{t_{m-2}} \frac{dt_{m-1}}{t_{m-1}}\int_0^{t_{m-1}} f(t_m)\,dt_m \\ \leftrightarrow\; \Psi_m(s) &= \int_s^\infty \frac{ds_1}{s_1}\int_{s_1}^\infty \frac{ds_2}{s_2}\int_{s_2}^\infty \ldots \int_{s_{m-1}}^\infty \frac{ds_m}{s_m} F(s_m)\,.\end{aligned} \tag{25}$$

For the value of $m$ for which we assume that $\psi_m(t) \to \psi_m(\infty)$, we have according to the classical final-value theorem that

$$\lim_{s\to 0}[s\Psi_m(s)] = \lim_{t\to\infty} \psi_m(t) = \psi_m(\infty)\,, \tag{26}$$

and our target here is just to express the left-hand side via $F(s)$.

Substituting $\Psi_m(s)$ from (25) to (26), we sequentially simplify the left-hand side of (26), using L'Hospital rule. The first step is:

$$\lim_{s\to 0}[s\Psi_m(s)] = \lim_{s\to 0}\Big[s\int_s^\infty \frac{ds_1}{s_1}\int_{s_1}^\infty \frac{ds_2}{s_2}\int_{s_2}^\infty \ldots \int_{s_{m-1}}^\infty \frac{ds_m}{s_m} F(s_m)\Big]$$

$$= \lim_{s\to 0}\frac{\displaystyle\int_s^\infty \frac{ds_1}{s_1}\int_{s_1}^\infty \frac{ds_2}{s_2}\int_{s_2}^\infty \ldots \int_{s_{m-1}}^\infty \frac{ds_m}{s_m} F(s_m)}{\dfrac{1}{s}}$$

$$= \lim_{s\to 0}\frac{\displaystyle -\frac{1}{s}\int_s^\infty \frac{ds_2}{s_2}\int_{s_2}^\infty \ldots \int_{s_{m-1}}^\infty \frac{ds_m}{s_m} F(s_m)}{-\dfrac{1}{s^2}}$$

$$= \lim_{s\to 0}\Big[s\int_s^\infty \frac{ds_2}{s_2}\int_{s_2}^\infty \ldots \int_{s_{m-1}}^\infty \frac{ds_m}{s_m} F(s_m)\Big]$$

which already includes not $m$, but only $m$-1 integrals. Writing again $s$ as $(1/s)^{-1}$ , and continuing with the L'Hospital rule in the same manner, we eliminate all of the integrals and come to the simple equality

$$\lim_{s\to 0}[s\Psi_m(s)] = \lim_{s\to 0}[sF(s)] \ . \tag{27}$$

In view of (26) and (27), (22) is proved:

$$\lim_{s\to 0}[sF(s)] = \psi_m(\infty) \ . \tag{28}$$

Since for all the functions $\psi_{m+n}(t)$, we have $\psi_{m+n}(\infty) = \psi_m(\infty)$, all these functions should have the same asymptotic behavior as $t \to \infty$, which concludes the proof of the theorem and states the "physical sense" of (18).

Equation (27) should be considered already for the classical case of $m = 0$.

## 7. A shift of the average value

In many potential applications, as, e.g., in queuing theory (see [2] and references there) only essentially positive functions, $f(t) > 0$, can be involved. Thus, we can suggest an up-shifting of functions that have alternating polarity. However, this is irrelevant for the unlimited functions like those of Section 2.

Theorem 2: *For the shift* $f(t) \to f(t) + K$, *with some* $K > 0$, $\psi_q(t) \to \psi_q(t) + K$, $\exists q$.

Proof: Obviously:

$$\frac{1}{t}\int_0^t [f(\lambda) + K]d\lambda = \psi_1(t) + K \ , \tag{29}$$

and then, similarly,

$$\frac{1}{t}\int_0^t [\psi_q(\lambda) + K]d\lambda = \psi_{q+1}(t) + K, \quad \exists q \ . \tag{30}$$

For the analytically important case of $f_p(t) = t^p \sin(t)$ no shift will make this function positive at infinity. The above remark that $f_p(t) = t^p \varphi(t)$ includes the case of the essentially positive functions as

$$f_p(t) = t^p \,|\sin t\,|, \quad f_p(t) = t^p \sin^2 t \ .$$

is thus very important and one notes also that in order to find applications for the multiple average, stochastic functions should be involved.